\documentclass{llncs}
\usepackage{graphicx}
\usepackage[fleqn]{amsmath}
\usepackage{amsfonts}
\usepackage[english]{babel}

\newcommand{\Real}{{\mathbb R}}

\numberwithin{subcase}{case}

\begin{document}

\title{An Upper Bound of $\frac{7}{6}$n for the Minimum Size 2EC on Cubic 3-Edge Connected Graphs\thanks{This research was partially supported by grants from the Natural Sciences and \mbox{Engineering} Research Council of Canada}}
 
\author{Philippe Legault}

\institute {School of Electrical Engineering and Computer Science (EECS), University of Ottawa
Ottawa, Ontario K1N 6N5, Canada\\\email{philippe@legault.cc}}

\maketitle

\begin{abstract}
In this paper, we study the minimum size 2-edge connected spanning subgraph problem (henceforth 2EC) and show that every \mbox{3-edge} connected cubic graph $G=(V, E)$, with $n=|V|$ allows a 2EC solution for $G$ of size at most $\frac{7n}{6}$, which improves upon Boyd, Iwata and Takazawa's guarantee of $\frac{6n}{5}$.
\end{abstract}

\section{Introduction}

Given an unweighted 2-edge connected graph $G=(V, E)$, the \emph{minimum size 2-edge connected spanning subgraph problem} (henceforth \emph{2EC}) consists of finding a 2-edge connected spanning subgraph of $G$ with the minimum number of edges. Multiple copies of an edge are not allowed, and are not necessary \cite{cheriyan}. The problem has important applications in the design of networks that can survive the failure of a link, but it is MAX SNP-hard even for special cases \cite{csaba}. 

Letting $x_{e}$ be a binary variable with value 1 when edge $e\in E$ is in a 2EC solution, 2EC can be formulated as an integer program as follows:\begin{equation}\label{ilpTwoEC}
     \begin{aligned}
      \text{Minimize}\quad & \sum_{e\in E}x_e&  \\
      \text{Subject to}\quad &\sum(x_{ij}:i\in S, j\notin S) \geq 2 &\text{for all } \emptyset \subset S \subset V,\\
       & 0 \leq x_e \leq 1\text{, and integer} &\text{for all } e \in E. \\
     \end{aligned}
\end{equation}

The linear programming (LP) relaxation of 2EC, denoted by 2EC$^\text{LP}$, is obtained by relaxing the integer requirement in (\ref{ilpTwoEC}). Let OPT$(G)$ and OPT$_{\text{LP}}(G)$ denote the optimal value of 2EC for $G$ and the value of its LP relaxation, respectively. In order to measure the quality of the LP relaxation as a lower bound, we define the \emph{unit integrality gap} $\alpha$2EC as the supremum of the ratio of OPT$(G)$ over OPT$_{\text{LP}}(G)$ over all input graphs $G$. This gives a measure of the quality of the lower bound provided by 2EC$^\text{LP}$, which is crucial for approximation algorithms and methods like branch and bound. Note that a polynomial-time constructive proof that the unit integrality gap equals $k\in\mathbb{R}$ would provide a $k$-approximation algorithm for 2EC.

\looseness=-1
While $\alpha$2EC has been intensely studied, little is known about it other than that $\frac{9}{8}\leq\alpha\text{2EC}\leq\frac{4}{3}$ \cite{vygen}. A natural next step is to focus on the simplest form of the problem known to remain NP-hard, when 2EC is restricted to graphs that are \emph{cubic} (i.e., every vertex has degree three) and 3-edge connected \cite{csaba}; we thus hope to generalizing successful methods. In this paper, we demonstrate an improved upper bound for $\alpha$2EC when restricted to 3-edge connected cubic graphs.

\section{Notation and Background}\label{notation}
This section is dedicated to basic notions of graph theory, which will be used throughout the paper. 

Let $G=(V, E)$ be a simple undirected graph with vertex set $V$, edge set $E$ and $n=|V|$. A graph is \emph{subcubic} if all vertices have degree 3 or less. A \emph{subgraph} of $G$ is a graph $H=(V_{H}, E_{H})$ where $V_{H}\subseteq V$ and $E_{H}\subseteq E$, such that if edge $uv\in E_{H}$, then $u, v\in V_{H}$. Subgraph $H$ is said to be \emph{spanning} if $V_{H}=V$. For any subgraph $H$ of $G$ we sometimes use the notation $E(H)$ and $V (H)$ to denote the edge set and the vertex set for $H$, respectively, and we use $\chi^{E(H)} \in\mathbb{R}^{E}$ to denote the incidence vector of subgraph $H$ (i.e. $\chi^{E(H)}_{e}$ is the number of copies of edge $e$ in $H$). A \emph{component} of $G$ is a subgraph of $G$ for which any two vertices are connected to each other by paths, and which is not connected to any other vertex in $G$.

For any subset $A\subseteq V$, the \emph{complement} of $A$ is $\overline{A}=V\setminus A$. For any two non-overlapping subsets $A, B\subseteq V$, the edge set between $A$ and $B$ is denoted by $E[A: B]$, i.e. $E[A: B] = \{ij\in E\ |\ i\in A, j\in B\}$. Note that the subset $F=\delta(A)$ of $E$ for some $A\subset V$ is called a \emph{cut} $\delta(A)=E[A:V\setminus A]$ as it disconnects $G$. Graph $G$ is considered $k$-edge connected if and only if all cuts of $G$ have size greater or equal to $k$, $k\in\mathbb{N}$. An edge cut that contains $k$ edges is a \emph{$k$-edge cut}. An edge cut $F$ of $G$ is \emph{essential} if $G'=(V, E\setminus F)$ has at least two components each containing more than one vertex. Each of those components is called a \emph{shore} of the cut. For a $k\in\mathbb{N}$, $G$ is said to be \emph{essentially $k$-edge connected} if and only if $G$ does not have an essential edge cut $F$ with $|F|<k$. If $G$ is cubic, simple and essentially 4-edge connected, then it is also 3-edge connected.

The function $|\delta(\cdot)|$ is \emph{symmetric submodular} for $G$, i.e. for every two sets $Y, Z\subseteq V$, the following two properties hold:
\begin{eqnarray}
|\delta(Y)|+|\delta(Z)|\geq |\delta(Y\cup Z)| + |\delta(Y\cap Z)|\text{,}\\
|\delta(Y)|+|\delta(Z)|\geq |\delta(Y\setminus Z)| + |\delta(Z\setminus Y)|\text{.}
\end{eqnarray}

The concept of convex combination is used extensively: in the context
of this paper, we say that a vector $y\in\mathbb{R}^{E}$ is a \emph{2EC convex combination} if there exist 2-edge connected spanning subgraphs $H_{i}$ with multipliers $\lambda_{i}\in\Real_{\geq 0}, i=1, 2, \dotsc, j$ such that $y=\sum_{i=1}^{j}\lambda_{i}\chi^{E(H_{i})}$ and $\sum_{i=1}^{j}\lambda_{i} = 1$.

\section{Contribution}

Our main result, which is proven in Section \ref{sectionProof}, is the following.

\begin{corollary}\label{mainCorollary}
Given a 3-edge connected cubic graph $G=(V, E)$ with $n=|V|$, there exists a 2-edge connected spanning subgraph with at most $\frac{7}{6}n$ edges.
\end{corollary}

\noindent In other words, we show that the unit integrality gap for 2EC is bounded above by $\frac{7}{6}$ for \mbox{3-edge} connected cubic graphs, which improves upon Boyd, Iwata and Takazawa's upper bound of $\frac{6}{5}$\,\cite{iwata}. Since 2EC restricted to 3-edge connected cubic graph is the simplest form of the problem known to remain NP-hard, it also implies that successful methods can be generalized. Our methods are not polynomial and thus, do not result in an approximation algorithm. Nevertheless, they give hope that a $\frac{7}{6}$-approximation algorithm exists, which would improve on the existing methods which gives a $\frac{6}{5}$-approximation \cite{iwata}. It would also extend Takazawa's $\frac{7}{6}$-approximation for bipartite cubic 3-edge connected graphs to all cubic 3-edge connected graphs \cite{takazawa}. In order to prove Corollary \ref{mainCorollary}, we prove the following stronger statement which uses the concept of convex combination.

\begin{theorem}\label{mainThm}
Given a 3-edge connected cubic graph \hbox{$G=(V, E)$}, the vector \mbox{$y\in\Real^{E}$} defined by $y_{e} = \frac{7}{9}$ for all $e\in E$ is a convex combination of incidence vectors of 2-edge connected spanning subgraphs $H_{i}$, $i=1, 2, \ldots, k$.
\end{theorem}

In Section \ref{sectionProof}, we also prove the following intermediary result for essentially 4-edge connected cubic graphs.
\begin{lemma} \label{submodularLemma}
Given an essentially 4-edge connected cubic simple graph $G = (V, E)$ with $|V| > 6$ and edges $au, uv, vc, vd \in E$, no two subsets $S, S'\subset V$ different than $\{u, v\}$ exist such that $au, vd \in \delta(S)$, $au, vc \in \delta(S')$, and $|\delta(S)|=|\delta(S')|=4$.
\end{lemma}

Besides being useful in our proof for the main result, Lemma\,\ref{submodularLemma} is also of independent interest because it is a new reduction operation that is very specific to essentially 4-edge connected cubic simple graphs. 

\section{Literature Review}

The first to improve the 2-approximation ratio for 2EC were Khuller and Vishkin \cite{khuller}, in 1994, with their $\frac{3}{2}$-approximation algorithm obtained using depth-first search trees and a method called ``tree-carving''. This was later refined in 1998 to $\frac{17}{12}$ by Cheriyan, Seb\H{o} and Szigeti \cite{cheriyan} via ear decompositions, then to $\frac{4}{3}$ by Seb\H{o} and Vygen \cite{vygen}, in 2014.  Krysta and Kumar \cite{kumar} improved the ratio to $\frac{4}{3}-\epsilon$ based on a charging scheme. Other results have been claimed, but have either been shown to be false, or are without conclusive proofs \cite{heeger}.

In parallel, much development also occurred for special cases, especially cubic and subcubic graphs. Krysta and Kumar \cite{kumar} designed a $\frac{21}{16}$-approximation algorithm for cubic graphs. For subcubic graphs, in 2002, Csaba, Karpinski and Krysta \cite{csaba} attained the approximation ratio of $\frac{5}{4} + \epsilon$, which was then improved to $\frac{5}{4}$ by Boyd, Fu and Sun \cite{fu} in 2014, using circulations. When 2EC is further restricted to 3-edge connected cubic graphs, we denote the work of Huh \cite{huh}, and of Boyd, Iwata and Takazawa \cite{iwata}, who achieved ratios of $\frac{5}{4}$, and $\frac{6}{5}$, respectively. Very recently, Takazawa\,\cite{takazawa} obtained a $\frac{7}{6}$-approximation algorithm for bipartite cubic 3-edge connected graphs, inspired from the work of Boyd, Iwata and Takazawa, by using 2-factors covering specified cuts.

\begin{figure}
\begin{center}
\includegraphics{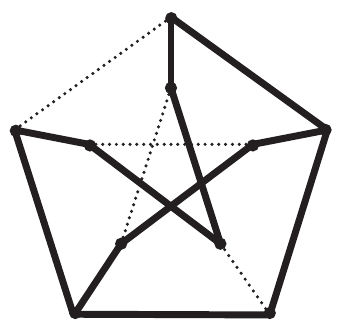}
\end{center}
\caption{The minimum size 2-edge connected spanning subgraph for the Petersen graph.}
\label{Petersen}
\end{figure}

In the general case, Seb\H{o} and Vygen proved that the unit integrality gap $\alpha$2EC is bounded as follows: $\frac{9}{8}\leq\alpha\text{2EC}\leq\frac{4}{3}$ \cite{vygen}. The integrality gap 2EC for cubic 3-edge connected graphs was shown in 2013 by Boyd, Iwata and Takazawa \cite{iwata} to be bounded above by $\frac{6}{5}$, and it is known that they are bounded below by the Petersen graph, at $\frac{11}{10}$, as shown in Figure \ref{Petersen} (where bold lines represent edges in the subgraph and dotted edges stand for edges in the Petersen graph, but not in the subgraph).

\section{2EC Convex Combinations for 3-edge Connected Graphs}\label{sectionProof}

In this section, we provide a proof for Lemma\,\ref{submodularLemma}, i.e. we show that for any essentially 4-edge connected cubic graph $G = (V, E)$ with $|V|>6$ and edges $au, uv, vc, vd \in E$, no two subsets $S, S'\subset V$ different than $\{u, v\}$ exist such that $au, vd \in \delta(S)$, $au, vc \in \delta(S')$, and $|\delta(S)|=|\delta(S')|=4$. This result will be used to prove the main theorem.

\subsection*{Proof of Lemma \ref{submodularLemma}.}

Here we provide a proof of Lemma \ref{submodularLemma}. Suppose that $G = (V, E)$ with $|V|>6$, edges $au, uv, vc, vd \in E$ and subsets $S, S'\subset V$ is the smallest counter-example to the lemma.

\begin{figure}
\begin{center}
\includegraphics{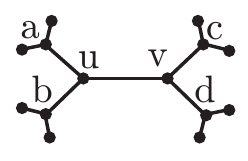}
\end{center}
\caption{Edges incident to $u$ and $v$ in $G$.}
\label{UvAndIncidentEdges}
\end{figure}

Let edges incident to $u$ and $v$ be as illustrated in Figure \ref{UvAndIncidentEdges}, with the unlabelled vertex neighbour to $u$ called $b$. Some edges incident to the unlabelled vertices are not shown.

\begin{figure}
\begin{center}
\includegraphics{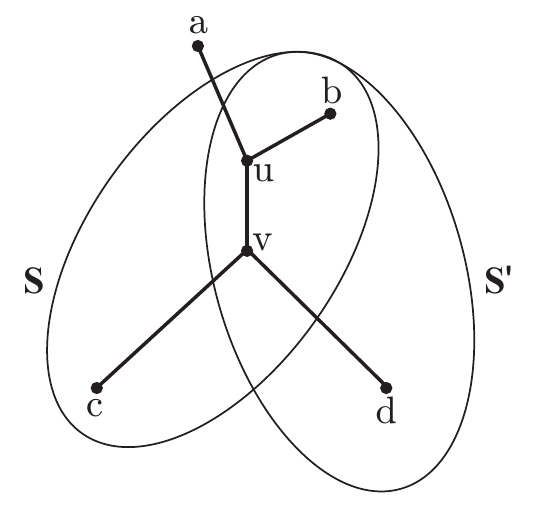}
\end{center}
\caption{Disposition of the vertices adjacent to $u$ and $v$ in $S$ and $S'$.}
\label{CutsDisposition}
\end{figure}

$G$ is simple and essentially 4-edge connected, and $|V|>6$, which implies that labelled vertices in Figure \ref{UvAndIncidentEdges} are distinct, as there would otherwise exist an essential 3-edge cut. Without loss of generality, assume that $u \in S$ and $u \in S'$ (if not, take the complement of the set). The cuts $\delta(S)$ and $\delta(S')$ are minimal and $G$ is cubic, which implies that two adjacent edges may not be in the same cut. Therefore $v, b, c \in S$ and $v, b, d \in S'$. We already know that $a, d \notin S$ and $a, c \notin S'$. Figure \ref{CutsDisposition} shows this disposition of the vertices in the cut.

Throughout this proof, we exploit the symmetric submodularity property of the function $|δ\delta(\cdot)|$, which states that
\begin{eqnarray}
|\delta(S)|+|\delta(S')|\geq |\delta(S\cup S')| + |\delta(S\cap S')|\text{,}\nonumber\\
8\geq |\delta(S\cup S')| + |\delta(S\cap S')|\text{.}\label{eqn1}
\end{eqnarray}

Because $u, v \in S \cap S'$, then $|S\cap S'| > 1$; similarly because $a \notin S\cup S'$, then
$\overline{|S \cup S'|} > 0$. It follows that $|\delta(S\cup S')| \geq 3$ and $|\delta(S\cap S')| \geq 4$ (since $G$ has no essential 3-edge cut). There exist two cases, when $|\delta(S\cup S')| = 3$ and $|\delta(S\cap S')|=4$ or $5$ and where $|\delta(S \cup S')| = 4$, because the cardinality of the cuts are restricted by (\ref{eqn1}).

\begin{case}\label{submodularCase1}
$|\delta(S\cup S')|=3$ and $|\delta(S\cap S')|=4$ or $5$.
\end{case}
Graph $G$ is simple and essentially 4-edge connected, which implies that one of the shores of $\delta(S\cup S')$ must consist of a single vertex, as the cardinality of this cut is 3: vertices $u$ and $v$ belong to $S\cup S'$, so $\overline{S\cup S'}=\{a\}$.

\begin{figure}
\begin{center}
\includegraphics{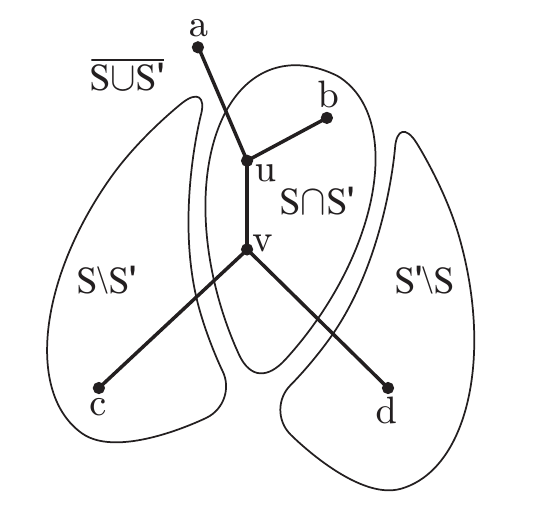}
\end{center}
\caption{Partition of $G$ to show the parity of parts, in Case \ref{submodularCase1}.}
\label{Partition}
\end{figure}

\looseness=-1
Any partition of a cubic graph has an even number of odd parts, because cubic graphs have an even number of vertices. We use this parity to show that $|\delta(S\cap S')| = 5$. The partition $\{S \setminus S', S' \setminus S, S \cap S', \overline{S \cup S'}\}$ (displayed in Figure \ref{Partition}) has $|\overline{S \cup S'}|$ odd, which implies that at least one of $S\setminus S'$, $S'\setminus S$ and $S\cap S'$ also has an odd number of vertices. The graph $G$ is cubic, which implies that for any subset $S\subseteq V$ of vertices, $\delta(S)$ is odd if and only if $|S|$ is odd. Given that $|\delta(S)| = |\delta(S')| = 4$, then $|S|$ and $|S'|$ must be even. This implies that the parity of $|S' \setminus S|$ and $|S \cap S'|$ must be the same; similarly, the parity of $|S \setminus S'|$ and $|S \cap S'|$ must also be the same. Therefore, either $|S \setminus S'|$, $|S' \setminus S|$ and $|S \cap S'|$ are all odd, or all even, to respect parity. But from the above, they cannot all be even. Thus, $|S \cap S'|$ is odd, which means that $|\delta(S \cap S')|$ must be odd as well. Therefore,
\begin{equation}
|\delta(S \cap S')| = 5\text{.}\label{SandSEquals5}
\end{equation}

All edges at $u$ and $v$ are known: let us examine an edge $bx$, for a vertex $x \in V$. We already know of 3 edges in $\delta(S \cap S')$, which means that there are exactly 2 others edges $e$ and $f$ in the cut. Assume that $x \in S \cap S'$: this means
that $bu$, $e$ and $f$ are in an essential 3-edge cut, because such a cut contains $b$ and $x$. Such a cut contradicts the fact that $G$ is essentially 4-edge connected. Therefore, $x \notin S \cap S'$, which immediately implies that $|S \cap S'| = 3$. We have that
\begin{eqnarray}
4=|\delta(S)| = |\delta(S\setminus S')| + |\delta(S\cap S')| - 2|E[S\setminus S': S\cap S'])|\text{,}\label{eqn5}\\
4=|\delta(S')| = |\delta(S'\setminus S)| + |\delta(S\cap S')| - 2|E[S'\setminus S: S\cap S']|\text{.}\label{eqn6}
\end{eqnarray}

\noindent We also have that $ba \notin E$ or else there would be an essential 3-edge cut
$\delta(\{u, b, a\})$, which means that from the five edges in $\delta(S \cap S')$, four link to either $S \setminus S'$ or $S' \setminus S$:
\begin{eqnarray}
4 = |E[S'\setminus S: S\cap S']| + |E[S\setminus S': S\cap S']|\text{.}\label{eqn7}
\end{eqnarray}

\noindent From the premise $|V | > 6$, it is implied that one of $S \setminus S'$ and $S' \setminus S$ contains more than one vertex. Without loss of generality, let it be $S \setminus S'$. Because $G$ has no essential 3-edge cut and $|S \setminus S'|>1$, we infer that.
\begin{eqnarray}
4 \leq |\delta(S \setminus S')| \text{,}\label{sNewBound}\\
3 \leq |\delta(S' \setminus S)|\text{.}\label{sPrimeNewBound}
\end{eqnarray}

\noindent The symmetric submodularity property ensures that
\begin{eqnarray}
|\delta(S)|+|\delta(S')|&\geq |\delta(S\setminus S')| + |\delta(S'\setminus S)|\nonumber\\
8&\geq |\delta(S\setminus S')| + |\delta(S'\setminus S)|\text{.}
\label{symMod2}
\end{eqnarray}

\noindent We will now use algebraic manipulations to show a contradiction. We subtract (\ref{eqn6}) from (\ref{eqn5}) and add (\ref{eqn7}) twice:
\begin{eqnarray}
8 = |\delta(S\setminus S')| - |\delta(S'\setminus S)| + 4|E[S'\setminus S: S\cap S']|\text{.}\label{eqn8}
\end{eqnarray}

\looseness=-1
\noindent In equation (\ref{eqn8}), $|\delta(S\setminus S')| - |\delta(S'\setminus S)|$ must be a multiple of 4. Because inequalities (\ref{sNewBound}), (\ref{sPrimeNewBound}) and (\ref{symMod2}) restrict the values of \mbox{$|\delta(S\setminus S')|$} and \mbox{$|\delta(S'\setminus S)|$}, their difference must be zero, i.e.
\begin{eqnarray}
|\delta(S\setminus S')|=4\text{.}\label{diffZero}
\end{eqnarray}
Equation (\ref{eqn8}) is simplified to
\begin{eqnarray}
2 = |E[S\setminus S': S\cap S']|\text{.}\label{eqn9}
\end{eqnarray}
We now conclude by substituting (\ref{SandSEquals5}), (\ref{diffZero}) and (\ref{eqn9}) in (\ref{eqn5}):
\begin{align}
& 4=|\delta(S\setminus S')| + |\delta(S\cup S')| - 2|E[S\setminus S': S\cap S']|\nonumber\\
& \quad = 4 + 5 - 2 \times 2\nonumber\\
& \quad = 5\nonumber\text{,}
\end{align}
which gives a contradiction.

\begin{figure}
\begin{center}
\includegraphics{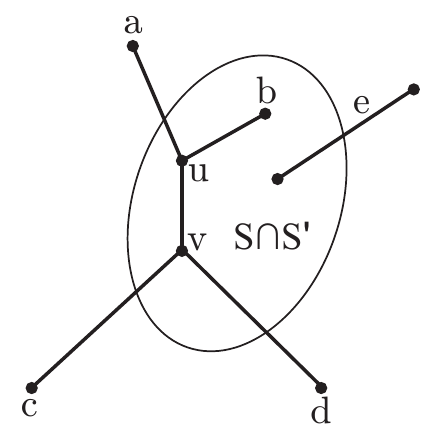}
\end{center}
\caption{Disposition of the vertices adjacent to $u$ and $v$ when $|\delta(S \setminus S')| = 4$.}
\label{Case2}
\end{figure}

\begin{case}
$|\delta(S \cup S')| = 4$ and $|\delta(S\cap S')|=4$.
\end{case}
Since $|\delta(S\cap S')|=4$, it follows that $|S\cap S'|$ is even, because $G$ is cubic. So there is at least one more vertex $w$ in $S\cap S'$. Since $G$ is 3-edge connected, at least one edge incident with $w$ is in $\delta(S\cap S')$. However, $vc, vd, au\in\delta(S\cap S')$ already, so we only have one other edge $e$ in the cut (as shown in Figure \ref{Case2}). Therefore $w$ is one end of $e$. This means that $\{ub, e\}$ is a 2-edge cut in $G$, which is a contradiction.\qed

\subsection*{Proof of Theorem \ref{mainThm}.}

\begin{figure}
\begin{center}
\includegraphics{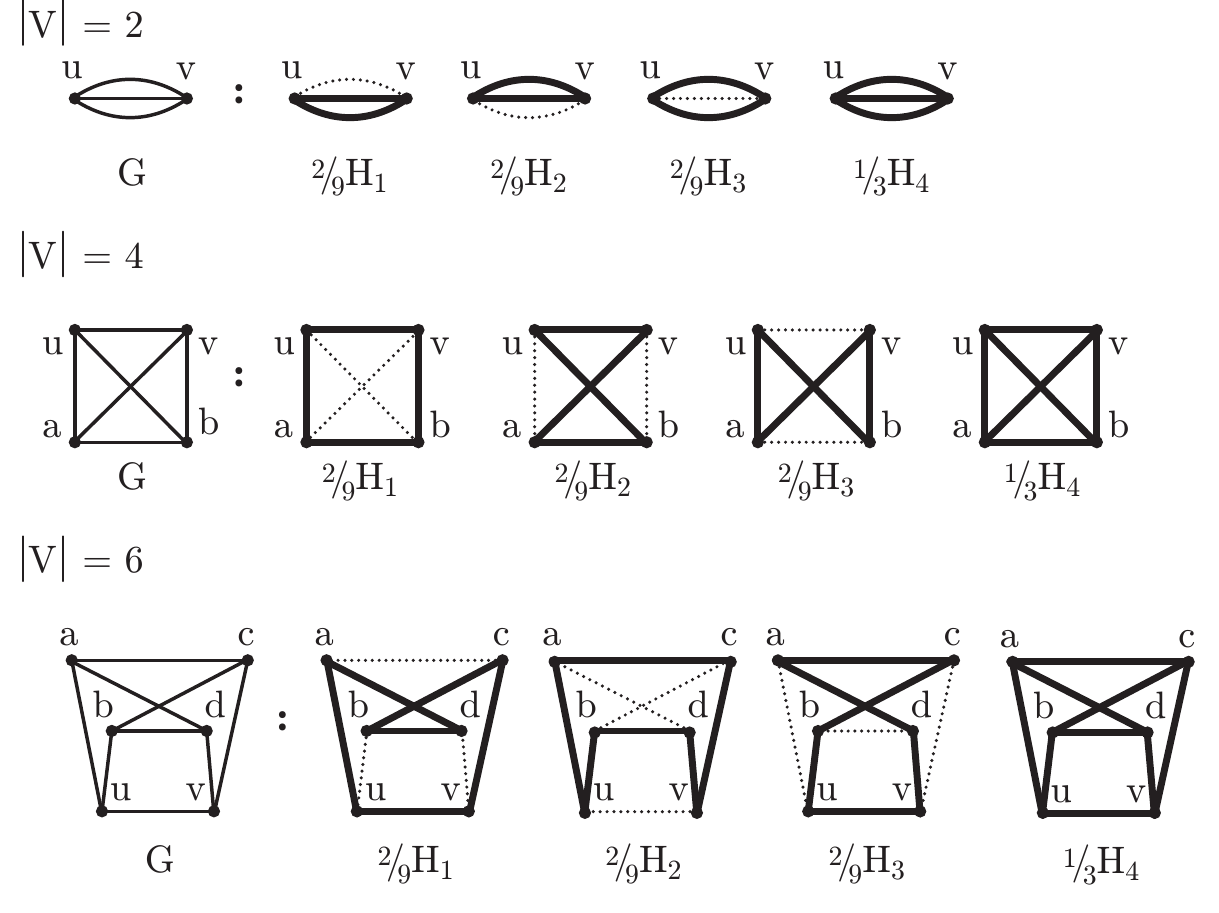}
\end{center}
\caption{Base cases for Theorem \ref{mainThm}. In the subgraphs $H_{i}$, bold lines indicate the edges in the subgraph and dotted lines indicate the edge omitted in the subgraph.}
\label{basecase}
\end{figure}

Let $G=(V, E)$ be the smallest counter-example for which the theorem does not hold. There are only three 3-edge connected essentially 4-edge connected cubic graphs with 6 vertices or less, on which the theorem can be shown to be true directly, as demonstrated in Figure \ref{basecase}, where bold lines represent edges in the subgraph, and dotted lines represent edges in $G$ not in the subgraph. In the figure, for each such graph $G$, the subgraphs $H_{i}$ and the corresponding $\lambda_{i}$ values for the required convex combination are shown. The smallest cubic 3-edge connected simple graph which is not essentially 4-edge connected has $|V|=6$, which means that either

\begin{enumerate}
\item $G$ is essentially 4-edge connected and $|V| > 6$,
\item $G$ has an essential 3-edge cut and $|V| \geq 6$.
\end{enumerate}

\setcounter{case}{0}
\begin{case}\label{case1}
$G$ is essentially 4-edge connected and $|V|>6$.
\end{case}
We pick an arbitrary edge $uv\in E$, and we label the other adjacent vertices of $u$ as $a$ and $b$, and the other adjacent vertices at $v$ as $c$ and $d$. We will use this edge to create the vector $z\in\mathbb{R}^{E}$ as a convex combination of incidence vectors of 2-edge connected spanning subgraphs, where for all $e\in E$, \[
  z_{e} =
  \begin{cases} 
      \hfill 1    \hfill & \text{ if $e = uv$,} \\
      \hfill \frac{1}{2}    \hfill & \text{ if $e \in \{ua, ub, vc, vd\}$,} \\
      \hfill \frac{8}{9}    \hfill & \text{ if $e \neq uv$ and $e$ adjacent to one of $ua, ub, vc$ or $vd$,} \\
      \hfill 1    \hfill & \text{ if $e \neq uv$ and $e$ adjacent to two of $ua, ub, vc$ or $vd$,} \\
      \hfill \frac{7}{9} \hfill & \text{ otherwise.} \\
  \end{cases}
\]
Graph $G$ has no essential 3-edge cut and $|V|>6$, which implies that $a$, $b$, $c$ and $d$ are distinct. Lemma \ref{submodularLemma} states that there does not exist two subsets $A, B\subset V$ different from $\{u, v\}$ such that $au, vc\in \delta(A)$, $au, vd\in \delta(B)$ and $|\delta(A)|=|\delta(B)|=4$. Thus, assuming that $\delta(B)$ is an essential 4-edge cut for some $B$ automatically means that $\delta(A)$ is not a 4-edge cut for any such $A$, and vice-versa. Without loss of generality, assume that edges $au$ and $vc$ are not in an essential 4-edge cut together, other than $\delta(\{u, v\})$, i.e. $\delta(A)>4$.

By Lemma \ref{submodularLemma} again, there does not exist two subsets $C, D\subset V$ different from $\{u, v\}$ such that $bu, vd\in \delta(C)$, $bu, vc\in \delta(D)$ and $\delta(C)=\delta(D)=4$. Knowing that $|\delta(A)| > 4$ for all $A$ such that $au, vc\in\delta(A)$ and $A\neq\{u,v\}$ implies that $|\delta(C)|>4$ for all $C$ such that $bu, vd\in\delta(C)$ and $C\neq\{u, v\}$, which means that edges $bu$ and $vd$ are not in an essential 4-edge cut together.

\begin{figure}
\begin{center}
\includegraphics{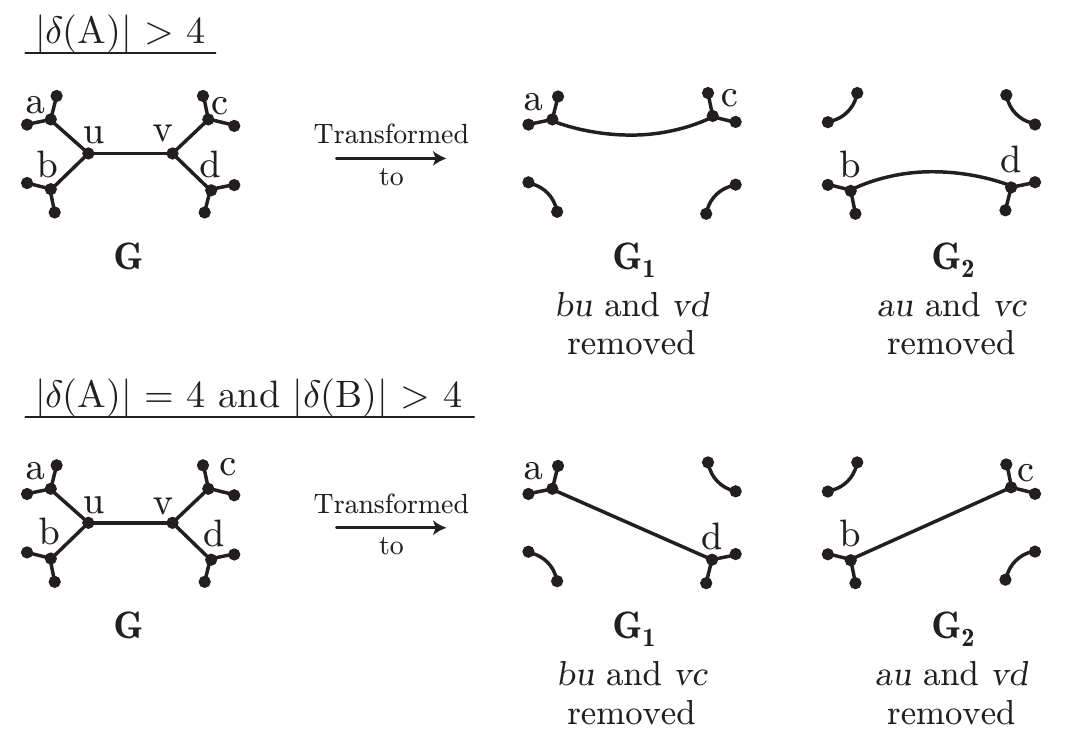}
\end{center}
\caption{Partial transformation of $G$ into $G_{1}$ or $G_{2}$ in Case 2 of Theorem \ref{mainThm}.}
\label{UVPartialTransform}
\end{figure}

Armed with these facts, we create graph $G_{1}$ by removing edges $au$ and $vc$, and $G_{2}$ by removing edges $bu$ and $vd$. In both graphs, any vertex $s$ of degree 2 with incident edges $st$ and $sr$ is removed, and the edge $sr$ is added. Because $\delta(A)>4$, $bd, ac\notin E$, no multi-edge are created. This situation is illustrated in the first part of Figure \ref{UVPartialTransform}; the second part of the figure illustrates the same situation, but when $|\delta(A)| = 4$ and $|\delta(B)| > 4$, for any $B$ such that $au, vd\in \delta(B)$ and $B\neq\{u, v\}$. 

The astute reader may ask what happends if an edge links one of $a$, $b$, $c$ or $d$: assume that such an edge $rs\in E$ exists, for $\{r, s\}\subset\{a, b, c, d\}$. It is immediately apparent that $r$ and $s$ cannot both be neighbor to the same vertex, either $u$ or $v$, otherwise there would exist an essential 3-edge cut, e.g. $\{rsu\}$. Without loss of generality, let $r$ be adjacent to $u$ and let $s$ be adjacent to $v$. For simplicity, let $r=a$ and $s=c$. Cut $\delta(\{a, c\})$ is an essential 4-edge cut and $ua, vc\in\delta(\{a, c\})$. By Lemma \ref{submodularLemma} and because $|V|>6$, edges $au$ and $vd$ are not in an essential 4-edge cut together, and edges $bu$ and $vc$ are not in an essential 4-edge cut together, which means that the transformation is valid.

Both $G_{1}$ and $G_{2}$ are 3-edge connected and cubic, because $G$ was simple and essentially 4-edge connected and the two edges removed in $G$ were not in an essential 4-edge cut together.

\begin{figure}
\begin{center}
\includegraphics{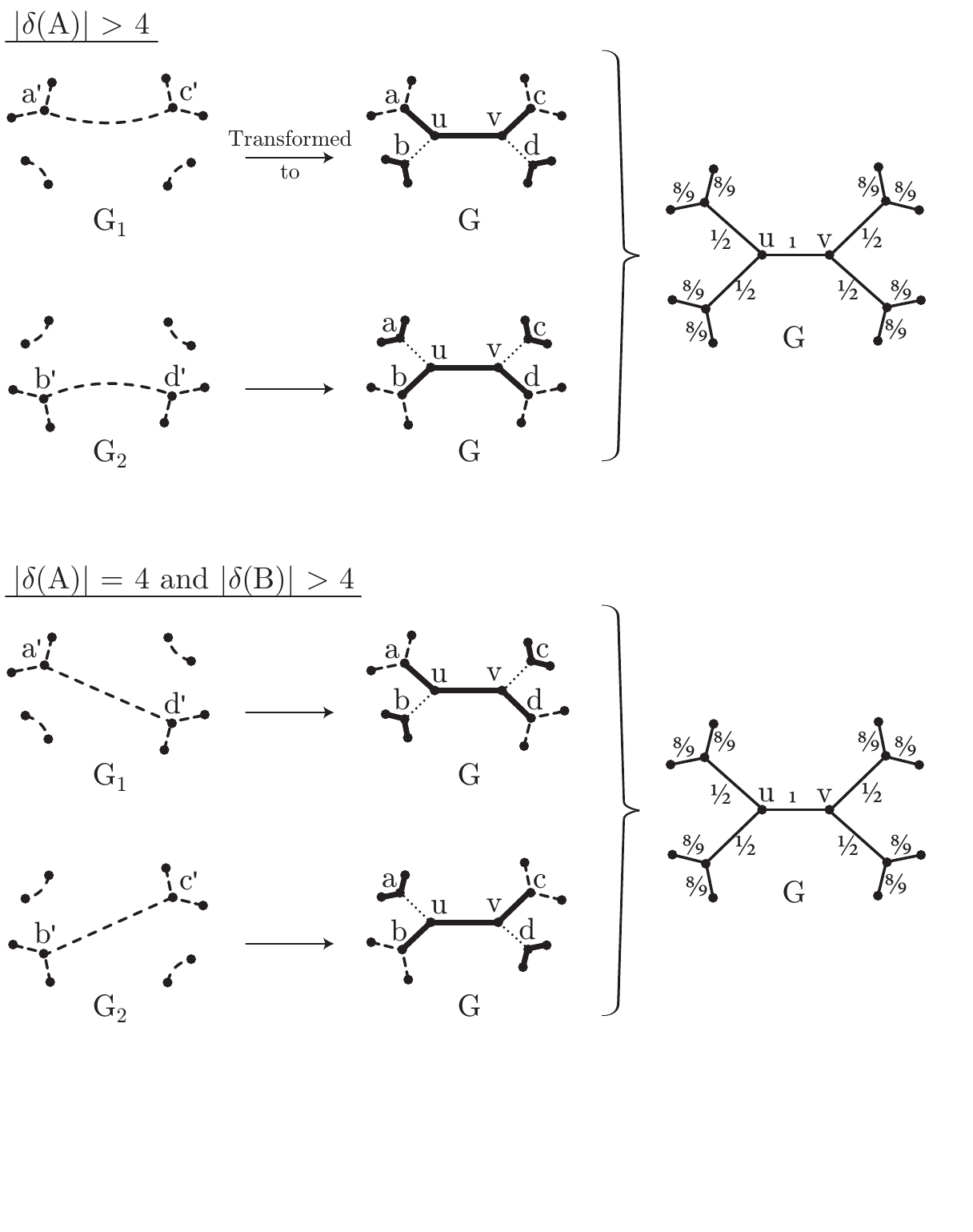}
\end{center}
\caption{Transformation of the 2-edge connected spanning subgraphs for $G_{1}$ and $G_{2}$ into 2-edge connected spanning subgraphs for $G$.}
\label{PatternsToMissingEdgeSevenNineth}
\end{figure}

The theorem holds for $G_{1}$ and $G_{2}$, as they are both smaller than $G$, and gives two convex combinations of incidence vectors of 2-edge connected spanning subgraphs $C_{1}$ and $C_{2}$ for $G_{1}$ and $G_{2}$, respectively. Note that an edge in the convex combination is present $\frac{7}{9}$ times overall in subgraphs. We will now modify the subgraphs in $C_{1}$ and $C_{2}$ to form sets $C_{1}'$ and $C_{2}'$ of subgraphs for $G$ in the following way: assume that edges $s$ and $t$ were the two edges that we removed to transform $G$ into $G_{i}$, then any 2-edge connected spanning subgraph for $G_{i}$ is 2-edge connected and spanning for $G$ with the same edge selection, save for edges adjacent to $s$ and $t$ which we always select, and $s$ and $t$ which we always omit. This is illustrated on the left side of Figure \ref{PatternsToMissingEdgeSevenNineth}, where bold lines represent edges in the subgraph, dotted line represent edges omitted in the subgraph and dashed lines represent edges which may or may not be in the subgraph. In the figure, edges incident to $a$, $b$, $c$ and $d$, but not adjacent to $uv$ are shown as distinct, for simplicity. 

We now take $\frac{1}{2}C_{1}'+\frac{1}{2}C_{2}'$ to obtain a 2EC convex combination for $G$. The occurrence of each edge in this convex combination is shown in the rightmost part of Figure \ref{PatternsToMissingEdgeSevenNineth}. Note that this is the occurrence for the case that no edges with both endpoints in the set $\{a, b, c, d\}$ exist. In the event such edges exist, then these edges have an occurrence of $1$, which does not affect the end result, as is shown later.

The convex combination $\frac{1}{2}C_{1}'+\frac{1}{2}C_{2}'$ is symmetrical (see the rightmost part of Figure \ref{PatternsToMissingEdgeSevenNineth}): therefore, repeating all the steps from Case \ref{case1}, taking every edge of $G$ in turn as edge $uv$ gives $|E|$ convex combinations which we label $F_{i}$, for $i=1,\ldots,|E|$. We build a new convex combination by setting the lambda value $\lambda_{F_{i}}=\frac{1}{|E|}$ for each $F_{i}$, and conclude that it selects every edge $\leq\frac{7}{9}$ times overall. We shall show this property in more detail here for an edge $yz\in E$: 
\begin{itemize}
\item edge $yz$ is treated as edge $uv$ in exactly one of the $|E|$ convex combinations. It then has an occurrence of 1.
\item there are no doubled edges, which means that $yz$ is adjacent to $uv$ in exactly 4 of the $|E|$ convex combinations. It then has an occurrence of $\frac{1}{2}$.
\item Because $G$ is cubic, there are exactly 8 ways to find an edge $yz$ exactly one edge away from $uv$, which are handled differently if $yz$ and $uv$ are in a 4-cycle together or not.
\subitem Assume that $yz$ and $uv$ are not in a 4-cycle $r$ times out of $|E|$, where $r \leq 8, r\in \mathbb{N}$ (the graph is cubic and simple, so there can be at most 8 such distinct edges for $yz$). It then has an occurrence of $\frac{8}{9}$.
\subitem If $yz$ and $uv$ are in a 4-cycle (i.e. $yu, zu\in E$), then $yz$ is one edge away from $uv$ both by $yu$ and $zu$. This happens $t$ times over the $|E|$ convex combinations and $yz$ then has an occurrence of 1. Because there are 8 ways to be one edge away from $uv$, and any 4-cycle with $yz$ and $uv$ uses two of those ways,
\begin{equation}
2t + r = 8,\text{ for }t\in \mathbb{N}\text{.} \label{firstLine}
\end{equation}
\item The occurrence of $yz$ in the other convex combinations is $\frac{7}{9}$.
\end{itemize}

The average occurrence of an edge over the $|E|$ convex combinations is 
\begin{align}
&\leq \frac{1}{|E|}\Big(1+t+4\times \frac{1}{2}+ r\times\frac{8}{9}+(|E|-r-t-5)\times\frac{7}{9}\Big)\nonumber\\
&=\frac{1}{|E|}\Big(\frac{2t}{9} + \frac{r}{9}-\frac{8}{9}+\frac{7|E|}{9}\Big)\nonumber\\
&= \frac{7}{9} + \frac{1}{9|E|}(2t+r-8)\text{.}\label{lastLine}
\end{align}

\noindent Since $2t + r = 8$, by (\ref{firstLine}), it follows from (\ref{lastLine}) that the average edge occurrence over the $|E|$ convex combinations is at most $\frac{7}{9}$. The convex combination which results from giving each convex combination a weight of $\frac{1}{|E|}$ selects every edge $\frac{7}{9}$ times. If an edge is selected less than $\frac{7}{9}$ of the time, we add it back arbitrarily to have it selected exactly $\frac{7}{9}$ of the time. The theorem holds.

\begin{case}
$G$ has an essential 3-edge cut $C$ and $|V| \geq 6$.
\end{case}

\begin{figure}
\begin{center}
\includegraphics{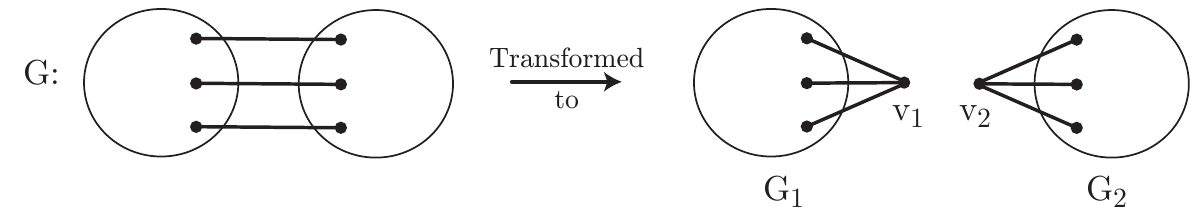}
\end{center}
\caption{Contracting both sides of an essential 3-edge cut of $G$.}
\label{3ECSplittingSevenNinth}
\end{figure}

Notice that the ends of the three edges in $C$ must be distinct because $G$ is \mbox{3-edge} connected. In this case we contract each shore of $C$ to a single pseudo-vertex, to obtain graphs $G_{1} = (V_{1},E_{1})$ with pseudo-vertex $v_{1}$ and $G_{2} = (V_{2},E_{2})$ with pseudo-vertex $v_{2}$ (as shown in Figure \ref{3ECSplittingSevenNinth}). Both $G_{1}$ and $G_{2}$ are smaller than $G$ so the theorem holds for $G_{1}$ and $G_{2}$. Moreover the patterns formed by the occurrence of the edges incident to $v_{1}$ and $v_{2}$ are unique and identical in the subgraphs in the corresponding convex combination. More specifically, exactly $\frac{2}{9}$ of the time, one of the incident edges will not be in the subgraph, on both sides of the cut, and this is true for each of the three incident edges. The remaining subgraphs contain all three incident edges. These constant patterns allow us to “glue” (reconnect the edges as there were before the inductive step) the subgraphs for $G_{1}$ and $G_{2}$ together, in such a way that identical patterns at $v_{1}$ and $v_{2}$ are matched and the resulting subgraphs are 2-edge connected. Figure \ref{3ECSplittingSevenNinth} displays the patterns formed by the occurrence of the edges incident to $v_{1}$ and $v_{2}$ and their occurrence, on the left. The figure also illustrates the process of gluing the 2-edge connected spanning subgraph in the convex combinations for $G_{1}$ and $G_{2}$ .

\begin{figure}
\begin{center}
\includegraphics{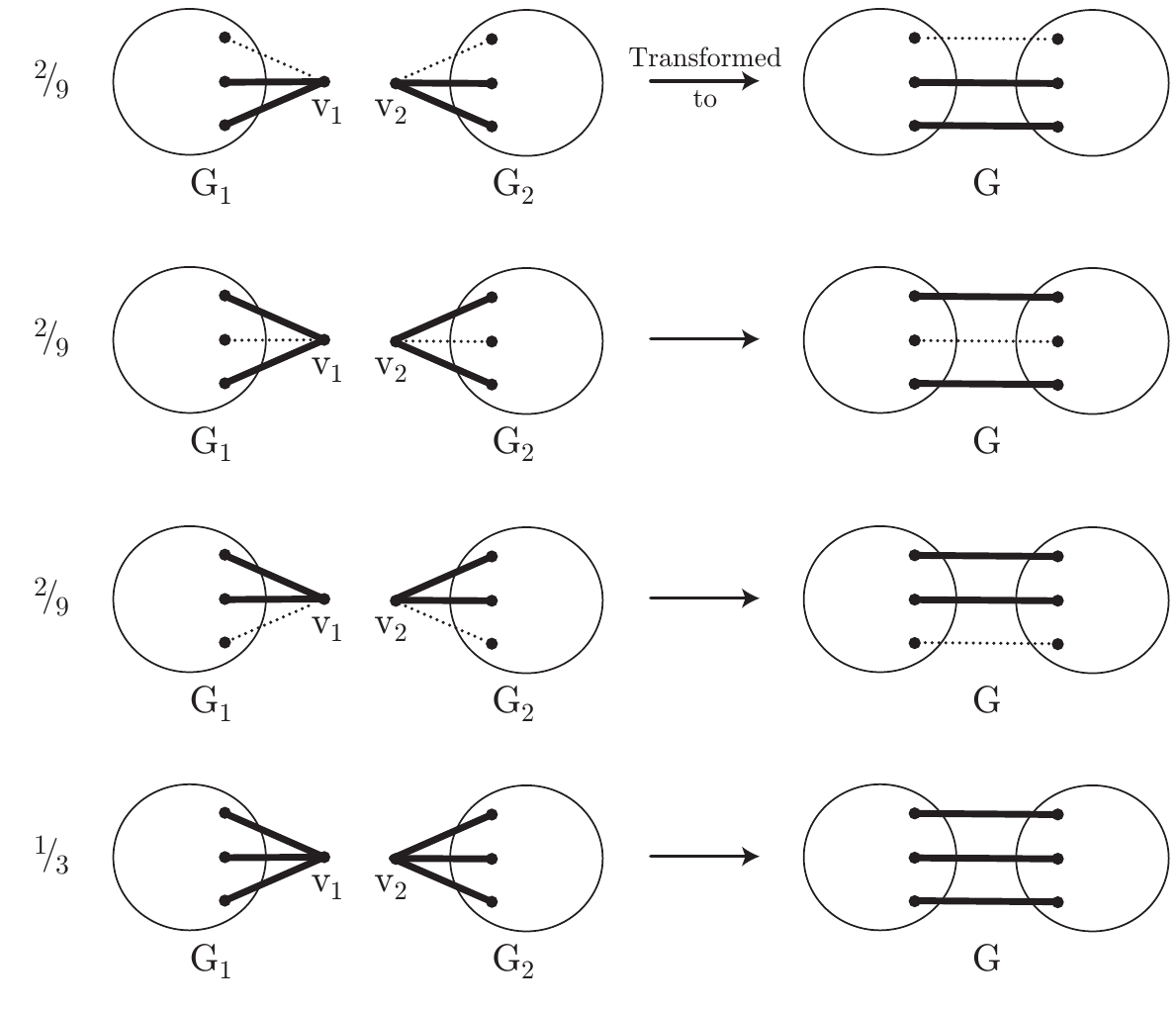}
\end{center}
\caption{``Gluing'' the 2-edge connected spanning subgraphs in the 2EC convex combinations for $G_{1}$ and $G{1}$.}
\label{3ECSplittingSevenNinth}
\end{figure}

This results in a convex combination that shows that the theorem holds for $G$, which gives a contradiction.\qed

\subsection*{Proof of Corollary \ref{mainCorollary}.}

\noindent Theorem\,\ref{mainThm} implies that for graph $G=(V, E)$, there exists a set of 2-edge connected spanning subgraphs $H_{i}$ with $\lambda_{i}\in\mathbb{R}_{\geq 0}$, $i=1,\ldots,j$ such that $\sum_{i = 1}^{j}\lambda_{i}=1$ and $\sum_{i = 1}^{j}\lambda_{i}\chi^{E(H_{i})}_{e}=\frac{7}{9}$ for all $e\in E$. This implies that for at least one of the $H_{i}$, $\sum_{e \in E}\chi^{E(H_{i})} \leq \frac{7|E|}{9}=\frac{7n}{6}$, since $|E|=\frac{3n}{2}$ for cubic graphs.\qed

\vspace{1em}

In other words, we show that the unit integrality gap for 2EC is bounded above by $\frac{7}{6}$ for \mbox{3-edge} connected cubic graphs, which improves upon Boyd, Iwata and Takazawa's upper bound of $\frac{6}{5}$ \cite{iwata}. Our methods are not polynomial and thus, do not result in an approximation algorithm. Nevertheless, they give hope that a $\frac{7}{6}$-approximation algorithm exists, which would improve upon the $\frac{6n}{5}$-approximation algorithm by Boyd, Iwata and Takazawa \cite{iwata}, and the $\frac{7n}{6}$-approximation algorithm when $G$ is bipartite, by Takazawa \cite{takazawa}.

\end{document}